# Evaluation of Energy-efficient VM Consolidation for Cloud Based Data Center - Revisited

Nasrin Akhter[1], Mohamed Othman[1], Ranesh Kumar Naha[2]

[1]Department of Communication Technology and Network, Universiti Putra Malaysia, 43400 UPM, Serdang, Selangor D.E., Malaysia.
[2]School of Engineering and ICT, University of Tasmania, Hobart, Australia.

nasrin786@gmail.com, mothman@upm.edu.my, raneshkumar.naha@utas.edu.au

**Abstract**
In this paper, a re-evaluation undertaken for dynamic VM consolidation problem and optimal online deterministic algorithms for the single VM migration in an experimental environment. We proceeded to focus on energy and performance trade-off by planet lab workload traces, which consists of a thousand Planetlab VMs with widespread simulation environments. All experiments are done in a simulated cloud environment by the CloudSim simulation tool. A new paradigm of utility-oriented IT services is cloud computing, which offers a "pay as you go" model. In recent years, there has been increasing interest among many users from business, scientific, engineering and educational territories in cloud computing. There is increasing concern that high energy consumption issues are a disadvantage for various institutions. However, so far too little attention has been given to the various methods to reduce energy consumption in cloud environments while ensuring performance. Besides the evaluation of energy-efficient data center management algorithms in the cloud, we proposed a further research direction toward the development of energy efficient algorithms. By the experimental evaluation of current proposal for competitive analysis of dynamic VM consolidation and optimal online deterministic algorithms for the single VM migration we found different results for different algorithm combinations. Cloud based data centers` consume massive energy, which has a negative effect on the environment and operational cost, this work contributes to the energy consumption reduction in cloud environment.

# 1 Introduction

Cloud computing offers computing services as a "pay-as-you-go" model, which enables utility computing, such as gas, electricity, water and telephone utilities. It could be referred to as the 5th utility[1]. Software as a Service (SaaS), Infrastructure as a Service(IaaS) and Platform as a Service (PaaS) are three commonly known cloud computing services. These services are deployed by following various deployment models. Depending on the requirements public, private, commodity and hybrid are four major cloud deployment models. Cloud computing helps users to save infrastructure and maintenance cost. Users receive scalability, reliability and mobile accessibility facilities from cloud computing services. Various companies focus on their innovation and creating business values by overlooking the low-level hardware and software infrastructure setup. The process of moving all computing services into the cloud is slowed down by various open challenges. Security, privacy, energy efficiency and common standards restrict many companies from migrating their computing services into the cloud. One of the most significant current discussions is energy saving issues. There is exceptional economic encouragement for data center operators, and it also supports greater environmental sustainability. A reduction in energy consumption, run quicker and cooler, and occupy less space are the most important challenges of modern large-scale computer systems like data centers and these contests are being dealt with concurrently by multicore processor technology. In the new global economy, high performance computing infrastructures have become a central issue to meet the demand for modern resource-intensive enterprises and scientific applications. The issue has grown in importance in light of the recent consumption of huge amounts of electrical power in large-scale computing data centers. However, these rapid changes are having a serious effect on the environment due to global energy consumption. The scale of data centers is increasing as is their operational cost. The predicted energy consumption cost will be 75% of the total cost as of 2014[2]. Energy and performance efficient algorithms for cloud based data centers presented in the paper[3]. In comparing optimal offline algorithms, they proved that optimal online deterministic algorithms considerably reduce energy consumption. We reviewed energy related issues in cloud data center in[23]. In this paper, we are going to implement the whole simulation environment proposed in the paper[3]. We used a combination of host overloading detection and VM selection algorithms for our simulation. Section

2 provides different aspects of energy savings in cloud computing. Section 3 begins with the description of the energy efficient algorithms for cloud based data centers. Section 4 elaborately states the experimental environment, simulation setup and performance metric. Section 5 analyzes the experimental results gathered from the simulations. Section 6 discusses some closely related works. Section 7 describes the work in progress.

Finally, the conclusion is given in the last section.

## 2 Energy Savings in Cloud Computing

Cloud computing used for the deployment of computing services in a business environment and also running many popular sites in the cloud. Reducing energy consumption is the most challenging research problem in the world of cloud computing. It is necessary to implement an energy efficient method for all system layers in order to reduce the energy consumption of a system. Services are running depending on the various demands of the users. Besides a faster and reliable service, users may be concerned about energy efficiency. A management system could be developed in order to track an energy-efficient service for future load distribution[4].

### 2.1 Energy-aware Data Centers

Nowadays virtualization is an important technology that enables energy-efficient services for servers in data centers. Putting idle servers in data centers in sleep mode is one of the ways to save energy. Every operative host produces heat. However, if we consolidate hardware and migrate to another underloaded host, then it is possible to reduce the energy consumption and deliver better energy efficiency. The cooling system of data centers is also responsible for incurring higher energy consumption costs. We may choose places with a lower temperature in a region to build a data center or develop new techniques for the cooling system of data centers for reducing power consumption.

## 2.2 Energy Savings in Networks and Protocols

A particular amount of energy consumed by network communication has been shown by research. The energy consumption of network communication must be improved by using QoS, performance and energy saving tradeoff [5]. The energy-efficient process of network elements may be increased if we find a way to redesign or optimize network protocols. The network devices are allocated to assign services to such devices that are energy efficient, and these devices are working while other devices are in sleep mode. In other words, network service allocation could be transferred from the energy inefficient devices to energy efficient devices. However, out-of-band signaling should be considered in order to develop protocols with energy awareness.

## 2.3 The effect of Internet applications

One large application region is information circulation in the Internet. However, at this time, most portions of Internet traffic are ruled by web based, video-on-demand, peer-to-peer and web services, and have consistently involved more than 85% of the Internet traffic for several years[6]. Internet traffic will increase more significantly when cloud computing becomes a major platform for creating and retrieving information. The ideal operation for replication of content and propagation algorithms will be energy, as the primary factor. Therefore, energy efficiency storage, computation, communication and performance, energy trade-offs should be reconsidered with respect to cloud computing.

## 3 Energy Efficient Algorithms

Several VMs could be run in a single machine based on a service request. In the virtualization concept, multiple operating systems could be run on multiple VM in a single physical machine. With VM consolidation, we can use system resources efficiently and unused resources could be placed in a low power state in order to save energy. Several heuristic methods have been proposed for VM consolidation in the paper[3]. For VM consolidation it is necessary to select the overloaded host first and we must follow some specific criteria for host overloading. After that, one or more hosts need to be migrated into an underloaded host or a new host. Depending on the load, several underloaded hosts could be migrated into one or more hosts. In this case, some underloaded hosts could be put in sleep mode after migrating the VMs running on them. Which VM and what type of VM could be migrated first is another consideration. Then, the selected VM will be placed into a new location. In the paper[3], the authors proposed five different host overload detection techniques and three different VM selection techniques for VM migration. Except Threshold (THR) VM overloading detection policy all other policies are based on statistical analysis. Median Absolute Deviation (MAD), Interquartile Range (IQR), Local Regression (LR) and Robust Local Regression (LRR) methods are used for detecting overloaded hosts. Minimum Migration Time (MMT), Random Choice (RC) and Maximum

Correlation (MC) policies are used for selecting VMs from overloaded hosts. In THR, 90% CPU utilization is the threshold for VM underload and overload detection. They also used Power Aware Best Fit Decreasing (PABFD) algorithm for VM placement. These algorithms find overloaded hosts first and migrate VMs until the hosts are considered overloaded. Then underloaded hosts will be selected based on utilization and some VM will be migrated in such a way that any host is not considered as overloaded.

## 4      Experimental and Simulation Setup

For our simulation, we used the CloudSim 3.0 toolkit[7], which is a modern simulation tool that supports the modelling of cloud data centers with on demand virtualization resources and application management. We modeled a data center with eight hundred heterogeneous physical machines and over 1000 running VMs in a simulation environment.

### 4.1      Workload, Power Model and DC Configuration

Each node of data center is modeled with one CPU core and the performance of each core is equivalent to 1000, 2000 or 3000 MIPS, 4 TB of storage and 32GB of RAM. The power consumption of hosts is defined according to the power consumption of HP Proliant G4 and G5 servers. According to the consumption of power of these servers, a server consumes from 175 W with 0% CPU utilization and maximum 250 W with 100% CPU utilization. Each VM requires one CPU core with 1000, 750, 500 or 250 MIPS, 1 GB of RAM and 100 GB of storage. Workload data has been taken from PlanetLab[8], which is a part of the CoMon[9] project. Random ten-day Workload traces collected from March 2011 to April 2011. CPU utilization was below 50% of the workload traces and during simulation VM assignment was random.

### 4.2  Physical Machine Configuration

We ran a simulation with an arrangement of three same type physical machines. The configuration of the physical machines is Intelr core™ 2 Duo CPU E8400 3.00 GHz processor and with 500GB storage, 4 GB RAM, Windows 7 with 32-bit OS and with 250 GB storage.

### 4.3  Performance Metrics

Performance is measured by following three main performance metrics. Total energy consumption: The total energy consumption is measured by total energy consumption by a data center during application workload. The unit for energy consumption is kilo watt per hour (kWh). SLA violation percentage: The SLA violation percentage is defined as the percentage of SLA violation events relative to the total number of the processed time frames. SLA violation is calculated through Performance Degradation due to Migration (PDM) and SLA Time per Active Host (SLATAH).
The number of VM migrations: The number of VM migrations initiated by the VM manager during the adaptation of the VM placement. Although, the objective of simulated algorithm is reducing violation of SLA and energy consumption, therefore, Energy and SLA Violation (ESV) is used as a combined metric.

## 5  Result Analysis

We implement back the benchmark work simulated in the CloudSim toolkit. The simulation architecture is shown in Figure 1.

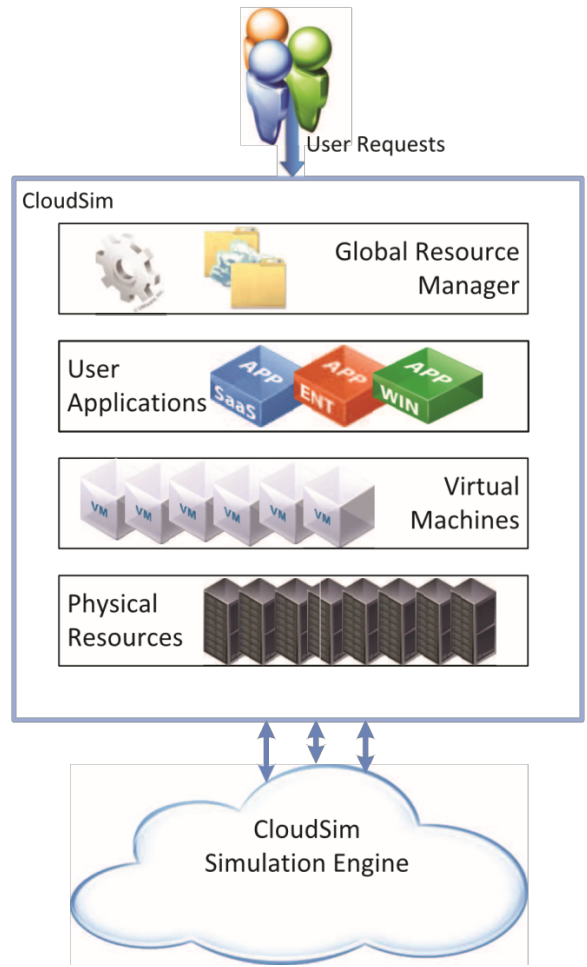

Fig. 1. Architecture of energy efficient algorithm simulation.

In our experiment, we used the collected data of CPU utilization from more than 500 places across the world for over one-thousand virtual machines in different servers. A utilization measurement of the interval is five minutes. We simulated all possible combinations for five hosts overloading detection and three VM selection algorithms. Various threshold combinations were used for different host overload detection algorithms, which resulted in 81 combinations. We conducted 810 simulations and summarize the best results from different algorithm combinations.

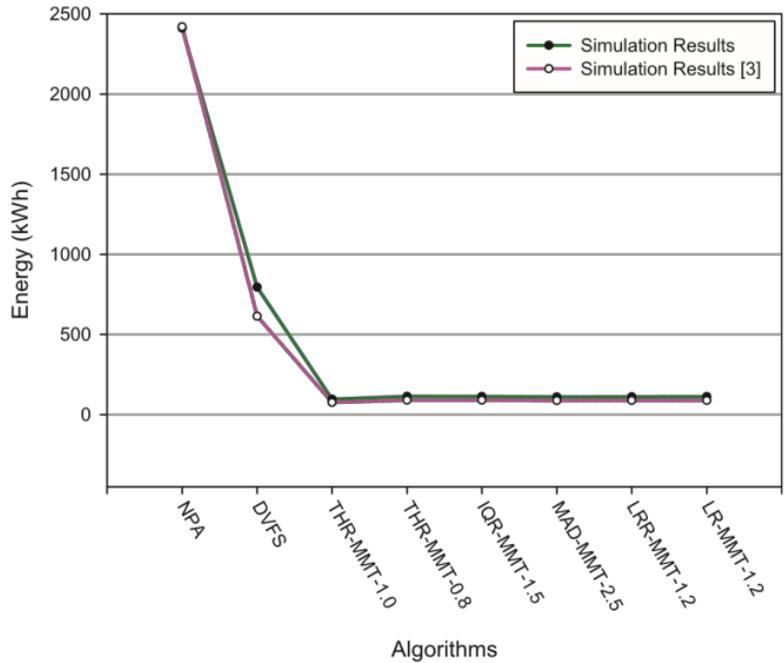

Fig. 2. Energy consumption comparison (median values).

From our simulation, we found that energy consumption slightly reduces for Non Power Aware (NPA) policy. However, using Dynamic Voltage Frequency Scaling (DVFS), VM overloading and selection policies produced results with increased energy consumption. Comparing previous research work, on average, we found 23% higher energy consumption from our simulation results. The simulation results for energy consumption are shown in Figure 2. SLA violation, VM Migration and ESV are '0' for NPA and DVFS, as shown in Figures 3, 4 and 5. NPA and DVFS do not use VMs consolidation and deconsolidation.

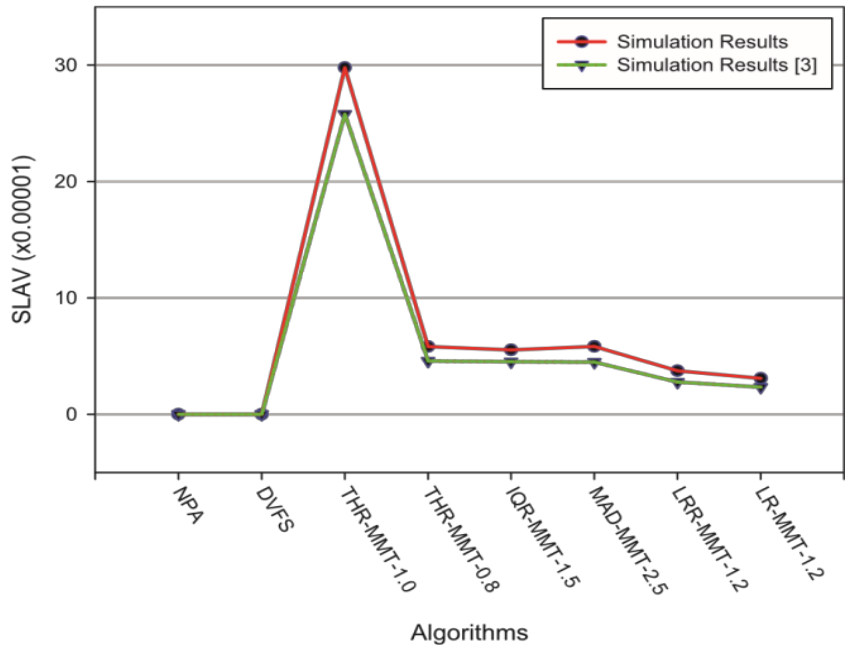

Fig. 3. SLA violation comparison (median values).

We found that the LR-MMT-1.2 algorithm shows 35% higher SLA violation compared to prior work. However, the average difference of SLA violation is 27% higher compared to prior work.

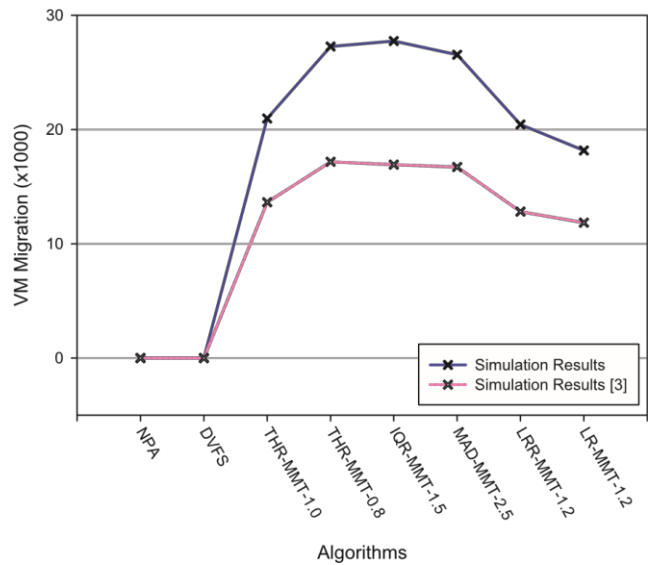

Fig. 4. SLA Comparison of VM migration (median values).

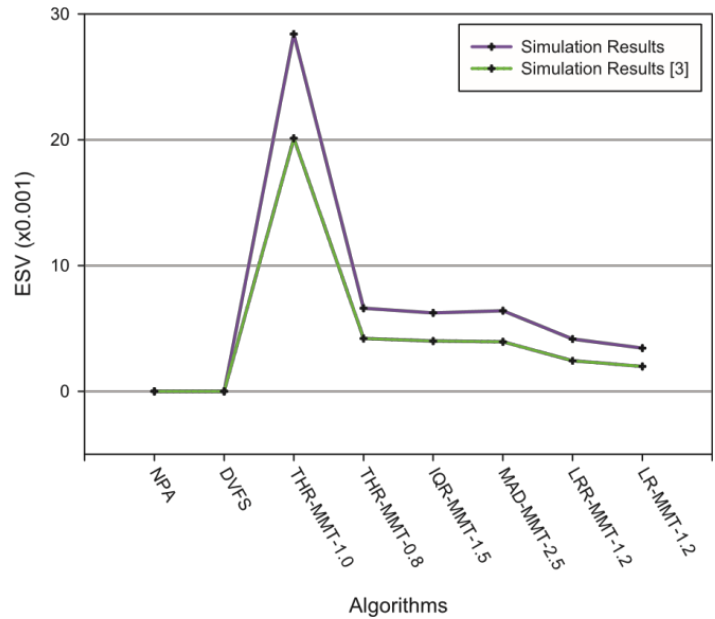

Fig. 5. Energy SLA violation comparison (median values).

About 60% higher VM migration and energy SLA violations were found in our simulation compared to prior work. Prior work was simulated using CloudSim 2.0. However, many features have been added in CloudSim 3.0. CloudSim3.0 has improved software requirements including Java version 1.6 or newer. Another point is that CloudSim3.0 has added 13 bugfixes, 7 new features and 17 major changes compared to version 2.0. Dynamic VM consolidation algorithms beat the static allocation policies, such as NPA and DVFS by showing significant energy efficiency. This means that the host switching to sleep mode using VM consolidation saves more energy. The work in[11], showed that approximately 450W power is consumed by a typical blade server in the totally utilized state while it consumes about 10.4 W power in sleep mode. During sleep mode, the transition delay is 300 ms.

## 6   Related Work

Power management in data centers was firstly applied by the paper[12]. In their work, the authors proposed a technique for the reduction of power consumption in a heterogeneous cluster in which computing nodes were used for various web applications. Depending on the workload, minimization of the physical nodes and the switching of idle nodes were the main techniques, which leads to a saving in power consumption. Dealing with power performance tradeoff is required for this approach, as performance degradation could occur during workload consolidation. A management system could be developed in order to track an energy-efficient service for future load distribution[10]. In a paper[13], authors described the problem of energy-efficient administration of uniform resources in Internet hosting centers. The main task is to define the resource requirement of each application and allocate resources in an efficient way considering the load level of the current request. Overall, these studies highlighted the management of the CPU without considering other system resources. In the work [15], authors studied the problem of power-efficient resource management in a single web-application environment with fixed SLAs (response time) and application handled load balancing. However, this management technique could be inefficient if the guest operating system is power-unaware or legacy. A large and growing body of literature explored the thermal-efficient resource management in data centers[14,16]. The studies showed that energy savings are possible by temperature-aware workload placement and software-driven thermal management. However, it is not a concern to analyze the problem of thermal management in the situation of virtualized data centers. In the paper[17], authors showed a competitive analysis on resource allocation and scheduling algorithms in cloud computing. From their analysis, they found that workflow scheduling maximizes profits and also advances execution efficiency without violating QoS requirements. In some papers[18,19,24] authors presented a simulation based evaluation of cloud brokering and load balancing algorithms without considering energy savings issues. Using Limited Lookahead Control (LLC), authors in the paper[20] defined the problem of power management in virtualized heterogeneous environments as a sequential optimization. The objective is to make best use of the resource provider profit by minimizing both SLA violation and power consumption. To estimate the number of future requests a Kalman filter is also applied to guess the future status of the system as well as perform the necessary reallocations. However, in contrast to heuristic-based methods, the suggested model requires simulation based learning for the application-specific adjustments, which cannot be implemented by IaaS Cloud providers, for example, Amazon EC2. Some literature addressed VM allocation and VM placement of cloud based data center[3,21]. These works demonstrated that dynamic workload usage in online deterministic algorithms could save significant energy consumption.

## 7   Work in Progress

Saving energy in cloud based data centers is a competitive research issue in order to reduce the carbon footprint, which also improves Return on Investment (ROI) of cloud service providers. Previous studies proved that VM consolidation and de-consolidation using various algorithms help to reduce energy consumption. These algorithms are based on many statistical methods, and by using these methods we can enable heuristic based analysis for VM consolidation and de-consolidation. These dynamic methods are energy efficient compared to static methods. In this work, we found different simulation results, which consume more energy as previous work claimed[3]. Following these results, presently we are continuing experiments and simulations using other basic statistical methods. We will customize these basic statistical methods in an energy efficient manner and apply in different experimental scenarios. Work on some new algorithms that can be used for energy efficiency in cloud environment is ongoing. It is not possible to implement our work in the real cloud, so we are going to use a real world complex workload for our simulation from PlanetLab. We will also compare with past deployed algorithms for VM allocation and placement. So far, we strongly believe that our algorithm will save more energy and also reduce SLA violations in cloud based data centers. Another direction of this research is to develop algorithms for energy savings, which will be distributive in nature and should be multi objective. Initial evaluation of our proposed algorithm was done in[22]. For energy efficient optimization, the optimization controller should be decentralized in order to avoid a single point of failure. A software platform is emerging for energy efficient resource allocation for cloud based data centers. Our ongoing works will provide new knowledge in the area of energy efficient resource allocation in cloud based data centers.

## 8   Conclusion

In this paper, we simulate existing energy efficient algorithms and come up with new findings. Our re-evaluation suggested that we might use some other statistical methods for saving energy in cloud based data centers. Cloud users are becoming more aware about environmental issues. The energy consumption of data centers is growing rapidly and produced higher energy consumption in the economic sector, which is a major source for $CO_2$ emissions. Many countries around the world are very concerned about energy policies in order to reduce greenhouse gas emissions. Further research is necessary to develop more energy efficient algorithms and also to focus on the existing architecture and develop a new software platform in order to support energy efficient functionalities.

**Acknowledgment**

This research is currently supported by the Malaysian Ministry of Education under the Fundamental Research Grant Scheme FRGS/1/11/SG/UPM/01/1.

**References**

[1] Buyya R, Yeo C S, Venugopal S, Broberg J, Brandic I. Cloud computing and emerging IT platforms: Vision, hype, and reality for delivering computing as the 5th utility. *Future Generation Computer Systems*, 2009, 25(6):599-616.

[2] Belady C. In the data center, power and cooling costs more than the it equipment it supports, 2007. [Online]. Available: http://www.electronicscooling.com/2007/02/in-the-data-centerpower-and-cooling-costs-more-than-theit-equipment-it-supports/.

[3] Beloglazov A, Buyya R. Optimal online deterministic algorithms and adaptive heuristics for energy and performance efficient dynamic consolidation of virtual machines in cloud data centers. *Concurrency and Computation: Practice and Experience*, 2012, 24(13):1397-1420.

[4] Moore J D, Chase J S, Ranganathan P, Sharma R K. Making Scheduling "Cool": Temperature-Aware Workload Placement in Data Centers. In *USENIX Annual Technical Conference, General Track.*, April 2005, pp.61-75.

[5] Gelenbe E, Silvestri S. Reducing power consumption in wired networks. In *Computer and Information Sciences, ISCIS 2009. 24th International Symposium on.*, September 2009, pp.292-297.

[6] Berl A, Gelenbe E, Di Girolamo M, Giuliani G, De Meer H, Dang M Q, Pentikousis K. Energy-efficient cloud computing. *The Computer Journal*, 2010, 53(7):10451051.

[7] Calheiros R N, Ranjan R, Beloglazov A, De Rose C A, Buyya R. CloudSim: a toolkit for modeling and simulation of cloud computing environments and evaluation of resource provisioning algorithms. *Software: Practice and Experience*, 2011, 41(1):23-50.

[8] Park k, Pai V S. CoMon: a mostly-scalable monitoring system for PlanetLab. *ACM SIGOPS Operating Systems Review*, 2006, 40(1):65-74.

[9] CoMon. 2014. [Online].   Available: http://comon.cs.princeton.edu/.

[10] Bash C, Forman G. Cool Job Allocation: Measuring the Power Savings of Placing Jobs at Cooling-Efficient Locations in the Data Center. In *USENIX Annual Technical Conference*, June 2007, pp.140.

[11] Meisner D, Gold B T, Wenisch T F. PowerNap: eliminating server idle power. In *ACM Sigplan Notices*, March 2009, pp.205-216.

[12] Pinheiro E, Bianchini R, Carrera E V, Heath T. Load balancing and unbalancing for power and performance in clusterbased systems. In *Workshop on compilers and operating systems for low power*, September 2001, pp.182-195.

[13] Chase J S, Anderson D C, Thakar P N, Vahdat A M, Doyle R P. Managing energy and server resources in hosting centers. In *ACM SIGOPS Operating Systems Review*, October 2001, pp.103-116.

[14] Tang Q, Gupta S K, Varsamopoulos G. Energy-efficient thermal-aware task scheduling for homogeneous highperformance computing data centers: A cyber-physical approach. *Parallel and Distributed Systems, IEEE Transactions on*, 2008, 19(11):1458-1472.

[15] Elnozahy E M, Kistler M, Rajamony R. Energy-efficient server clusters. In *PowerAware Computer Systems*, Springer Berlin Heidelberg, 2003, pp.179-197.

[16] Sharma R K, Bash C E, Patel C D, Friedrich R J, Chase J S. Balance of power: Dynamic thermal management for internet data centers. *Internet Computing, IEEE*, 2005, 9(1):42-49.

[17] Ma T, Chu Y, Zhao L, Ankhbayar O. Resource Allocation and Scheduling in Cloud Computing: Policy and Algorithm. *IETE Technical Review*, 2014, 31(1):4-16.

[18] Naha R K, Othman M. Brokering and Load-Balancing Mechanism in the Cloud – Revisited. *IETE Technical Review*, 2014, 31(4):271-276.

[19] Naha R K, Othman M, Optimized Load Balancing for Efficient Resource Provisioning in the Cloud. In *International Symposium on Telecommunication Technologies (ISTT), IEEE*, November 2014, pp.382-285.

[20] Kusic D, Kephart J O, Hanson J E, Kandasamy N, Jiang G. Power and performance management of virtualized computing environments via lookahead control.*Cluster computing*, 2009, 12(1):1-15.

[21] Beloglazov A, Abawajy J, Buyya R.Energy-aware resource allocation heuristics for efficient management of data centers for cloud computing. *Future Generation Computer Systems*, 2012, 28(5):755768.

[22] Akhter N, Othman M. Energy Efficient Virtual Machine Provisioning in Cloud Data Centers. In *International Symposium on Telecommunication Technologies(ISTT), IEEE*, November 2014, pp.282-286.

[23] Akhter N, Othman M. Energy aware resource allocation of cloud data center: review and open issues. In *Cluster computing*, 2016, 19(3):1163-1182.

[24] Naha R K, Othman M. Cost-aware service brokering and performance sentient load balancing algorithms in the cloud. *Journal of Network and Computer Applications*, 2016, 75:47-57.